\documentclass[floatfix,reprint,amsmath,amssymb,aps,pra]{revtex4-1}
\usepackage{times,mathptmx}
\usepackage{graphicx}
\usepackage[english]{babel}
\usepackage{epstopdf}
\usepackage{longtable}
\usepackage{xcolor}

\begin{document}

\title{Tunable virtual gain in resonantly absorbing media}
\author{Denis~V.~Novitsky}
\email{dvnovitsky@gmail.com}
\affiliation{B.~I.~Stepanov Institute of Physics, National Academy of Sciences of Belarus, Nezavisimosti Avenue 68, 220072 Minsk, Belarus}

\date{\today}

\begin{abstract}
Virtual gain refers to the simulation of real light amplification using radiation with exponentially decaying amplitude, so that its complex frequency corresponds to the scattering pole. We theoretically study virtual gain in a two-level resonant medium for different regimes of light-matter interaction depending on the radiation intensity. We show that virtual gain at the pole can be most clearly observed for low intensities, when the medium is absorbing, in contrast to the saturated medium at high intensities. The efficiency of virtual gain can be tuned with the light intensity and can be controlled dynamically through the population inversion of the medium. Our results show that resonantly absorbing media paradoxically mimics gain-like response, which admit a number of related phenomena and methods to mold both optical signals and material properties without relying on instability-prone gain media.
\end{abstract}

\maketitle

\section{Introduction}

Recent years have evidenced the rise of nanophotonics studying the methods for controlling light with small-scale (nanostructured) objects \cite{GaponenkoBook}. This approach has breathed new life into such classic subjects of optics as scattering. In particular, subtle tuning of system parameters has made possible to observe different anomalies in light scattering \cite{Krasnok2019}. These anomalies can be often associated with scattering poles and zeros, which can be conveniently represented on the complex-frequency plane. Generally, the poles and zeroes appear at the complex frequencies, but their position can be regulated by introducing gain or loss, thus, making the system non-Hermitian. When a pole or zero goes to the real-frequency axis, one observes lasing or antilasing (coherent perfect absorption, or CPA \cite{Chong2010}), respectively. Specific non-Hermitian systems obeying parity-time ($\mathcal{PT}$) symmetry demonstrate both lasing and antilasing simultaneously \cite{Longhi2010, Wong2016}. This is the so-called CPA-lasing effect obtained under the coalescence of a pole and a zero at the real-frequency axis. Analogous coalescence in a Hermitian system corresponds to a bound state in the continuum (BIC) -- nonradiating mode having infinite quality factor despite the system openness \cite{Hsu2016}. Sometimes, one can transform the BIC into CPA-lasing point by using geometric or non-Hermitian perturbations \cite{Novitsky2022}. Among other scattering anomalies, we mention exceptional points (non-Hermitian degeneracies) \cite{Miri2019}, anapoles \cite{Baryshnikova2019}, and superscattering \cite{Tribelsky2006, Ruan2010}.

Recently, dynamic effects have come under the spotlight strongly widening the range of optical systems and applications under study. In particular, materials with time-varying parameters (such as refractive index) attract much attention \cite{Galiffi2022} as a counterpart to the familiar space-varying systems such as interfaces \cite{Xiao2014, Plansinis2015}, layers \cite{Ramaccia2020}, photonic crystals \cite{Biancalana2007}, metamaterials \cite{Yin2022}, and metasurfaces \cite{Mostafa2022}. A number of temporal analogs of usual optical phenomena were reported, including Brewster angle \cite{Pacheco2021}, Anderson localization \cite{Sharabi2021}, topological protection \cite{Lustig2018}, $\mathcal{PT}$ symmetry \cite{Li2021}, negative refraction \cite{Lasri2022}, etc. What is most important for our discussion, variation of media in time serves as a source of energy flowing into the system and allowing to circumvent the law of energy conservation. As a result, frequency and momentum switch their roles making possible such effects as the frequency conversion under time modulation of materials \cite{Morgenthaler1958}. In the temporal analogs of photonic crystals, one can observe the momentum bandgaps inside which one can observe amplification of waves \cite{Holberg1966, Lyubarov2022, Sharabi2022, Wang2022}. However, practical possibilities of time-varying photonics are strongly limited by the difficulties with fast and spatially uniform modulations of materials and by interference with common nonlinear effects \cite{Hayran2022}.

Another possibility is to harness time variation not for the medium, but for light signal itself. As a result, not medium, but radiation should be described with the complex frequency making it ``non-Hermitian''. This possibility was implemented in the idea of virtual loss and gain, which attracted a lot of attention recently. Virtual loss (or virtual perfect absorption, VPA) can be considered as an imitation of CPA by taking radiation with exponentially growing amplitude. If the rate of exponential growth (i.e., the imaginary part of complex frequency of radiation) is equal to the imaginary part of scattering zero frequency, radiation seems to be absorbed \cite{Baranov2017}. In fact, it is stored inside the medium and released after the exponential signal is switched off. Theoretical idea of virtual loss was initially demonstrated with the examples of dielectric slab and cylinder \cite{Baranov2017}. Subsequently, it was expanded to the discrete array of resonators or waveguides \cite{Longhi2018}, ring microcavity coupled to a waveguide \cite{Zhong2020}, and open metasurface-based cavity \cite{Marini2020}. The effect was experimentally proved with elastic waves \cite{Trainiti2019}. Virtual gain is a counterpart of virtual loss engaging a scattering pole under exponentially decaying irradiation \cite{Li2020}. As further generalizations, we mention the virtual $\mathcal{PT}$ symmetry with balanced loss and gain \cite{Li2020} and the certain radiation waveforms absorbed at the exceptional points \cite{Farhi2022}. The virtual processes based on manipulations with the time-varying light signals are finding such applications as critical coupling for transformation of electromagnetic energy \cite{Radi2020}, virtual pulling forces for particles transportation \cite{Lepeshov2020}, enhancement of light scattering \cite{Ali2021, Kim2022}, and topological light localization via non-Hermitian skin effect \cite{Gu2022}. It should be emphasized that the total energy of radiation does not grow under virtual gain, so that there is no risk of instability usual for systems with light amplification (such as $\mathcal{PT}$-symmetric ones).

\begin{figure}[t!]
\includegraphics[scale=1.5, clip=]{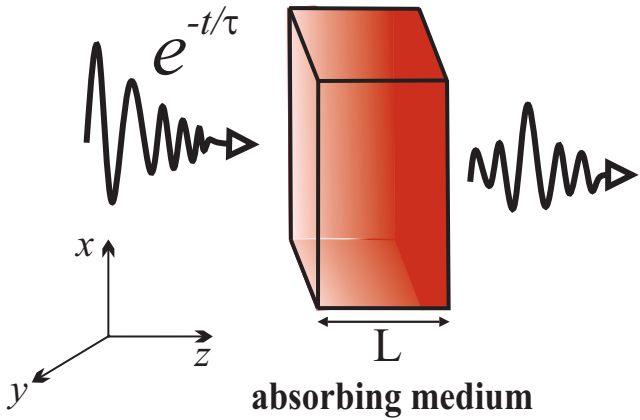}
\caption{\label{fig1} Schematic representation of the situation studied: An exponentially decaying radiation is impinging on the layer of resonantly absorbing medium. The color gradient inside the medium shows the growing intensity of light corresponding to virtual gain.}
\end{figure}

In this paper, we apply the idea of virtual gain to the resonantly absorbing media. The two-level model of resonant quantum media have attracted much attention since the beginning of laser era \cite{AllenBook}. Pulsed-field dynamics in such media was studied in much detail on the basis of Maxwell-Bloch equations \cite{Lamb1971, Kryukov1970}. In particular, coherent pulses (much shorter than relaxation times) evidently violate Beer's law propagating almost without attenuation due to such phenomena as self-induced transparency \cite{McCall1969, Poluektov1975} and zero-area pulse formation \cite{Crisp1970}. The further advances include studies of the ever more shorter -- few-cycle \cite{Ziolkowski1995, Kalosha1999, Tarasishin2001}, subcycle \cite{Novitsky2012}, and unipolar \cite{Arkhipov2020} -- pulses as well as interactions between pulses in resonant media \cite{Afanas'ev1990, Shaw1991} resulting in such phenomena as population gratings \cite{Arkhipov2021} and diode-like effect \cite{Novitsky2012a}, among others. On the other hand, the incoherent waveforms with longer characteristic timescales such as incoherent solitons \cite{Afanas'ev2002} and optical kinks \cite{Ponomarenko2010, Novitsky2017} can also withstand absorption in resonant media. We also mention the studies of ultrashort pulses \cite{Novitsky2018} and wavefront propagation \cite{Novitsky2021} in disordered resonantly absorbing and amplifying media. The role of driving-field decay rate in superradiance and subradiance dynamics was revealed recently \cite{Asselie2022}. However, the virtual-gain dynamics of quasi-continuous patterns without evident propagation effects connected with pulses or wavefronts have not been studied earlier.

Using numerical simulations of quasi-continuous monochromatic and then exponentially-decaying radiation in a two-level medium, we analyze the different regimes depending on light intensity. In the low-intensity regime, we have the classic virtual gain, when the signal attenuating inside the medium is replaced by the gradually growing waveform. In the high-intensity regime, the medium gets saturated with the step-like patterns seen in the system's response. Moreover, one can switch between these different types of response by simply choosing the time instant to start the signal decay. Our results clearly point to the connection of the virtual gain effect with the position of pole in the complex-frequency plane and can be used for controlling both the radiation patterns and quantum-medium state with time-varying light signals.

\section{Resonant medium description}

We consider the two-level resonant medium of thickness $L$ illuminated by the normally incident radiation of wavelength $\lambda$ (Fig. \ref{fig1}). Interaction of light with the medium is described with the well-known Bloch equations \cite{Kalosha1999},
\begin{eqnarray}
\frac{d \rho_{12}}{d t} &=& i \omega_0 \rho_{12} + i
\frac{\mu}{\hbar} E w - \frac{\rho_{12}}{T_2}, \label{polar}
\end{eqnarray}
\begin{eqnarray}
\frac{d w}{d t} &=& -4 \frac{\mu}{\hbar} E \textrm{Im} \rho_{12} -
\frac{w+1}{T_1}, \label{invers}
\end{eqnarray}
where $E$ is the electric field, $\rho_{12}$ is the off-diagonal density matrix element (atomic polarization), $w=\rho_{22}-\rho_{11}$ is the population inversion (difference between populations of the excited and ground levels), $\omega_0$ is the resonance frequency, $\mu$ is the dipole moment corresponding to the transition between levels, $T_1$ and $T_2$ are the relaxation times, $c$ is the speed of light, and $\hbar$ is the Planck constant. These equations should be completed with the wave equation accounting for light propagation,
\begin{eqnarray}
\frac{\partial^2 E}{\partial z^2}&-&\frac{n^2}{c^2} \frac{\partial^2
E}{\partial t^2} = \frac{4 \pi}{c^2} \frac{\partial^2 P}{\partial
t^2}, \label{Max}
\end{eqnarray}
where $P=2 \mu C \textrm{Re} (\rho_{12})$ is the macroscopic polarization, $n$ is the background refractive index, and $C$ is the concentration of two-level atoms. Note that these equations do not employ the popular rotating-wave approximation (RWA) and slowly-varying-envelope approximation (SVEA). The reason is that the fast change of light field can break these approximations \cite{Ziolkowski1995, Kalosha1999, Tarasishin2001, Novitsky2012}. We address the applicability of RWA in Appendix \ref{RWA} with the conclusion of qualitative agreement between the general and RWA cases. The Maxwell-Bloch equations (\ref{polar})-(\ref{Max}) are numerically solved using the approach described earlier \cite{Novitsky2009}. Further, we adopt the parameters of the medium as follows: the thickness $L=\lambda=1$ $\mu$m; the background refractive index $n=3$; the relaxation rates $T_1=1$ ns and $T_2=1$ ps; the density parameter governing the strength of light-matter interaction $g=4 \pi \mu^2 C/3 \hbar=3.7 \cdot 10^{11}$ s$^{-1}$; these values can be reached, e.g., for quantum-dot media. Further, we also assume that radiation is tuned exactly to the resonance with the quantum transition, i.e., $\omega=2 \pi c/\lambda=\omega_0$ with $\lambda=1$ $\mu$m.

\begin{figure}[t!]
\includegraphics[scale=1., clip=]{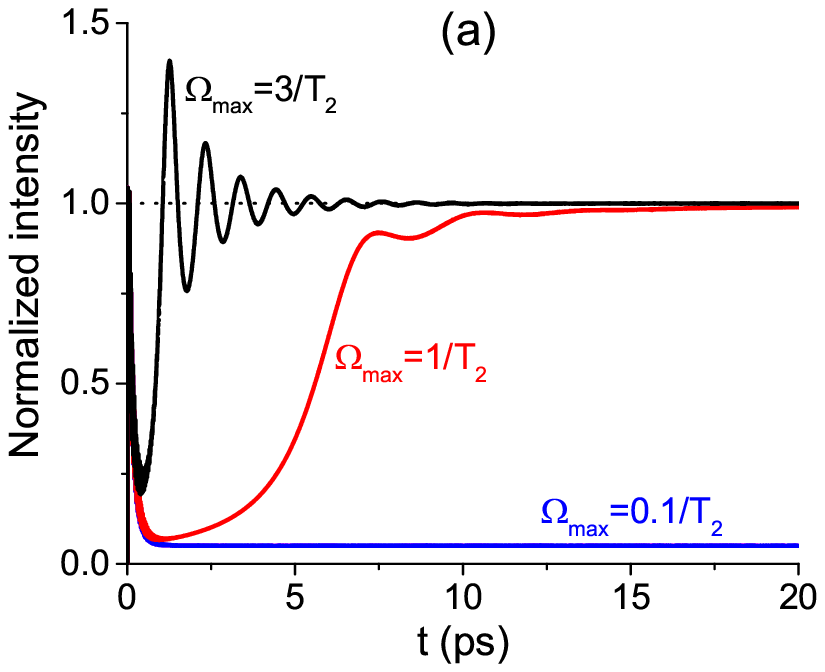}
\includegraphics[scale=1., clip=]{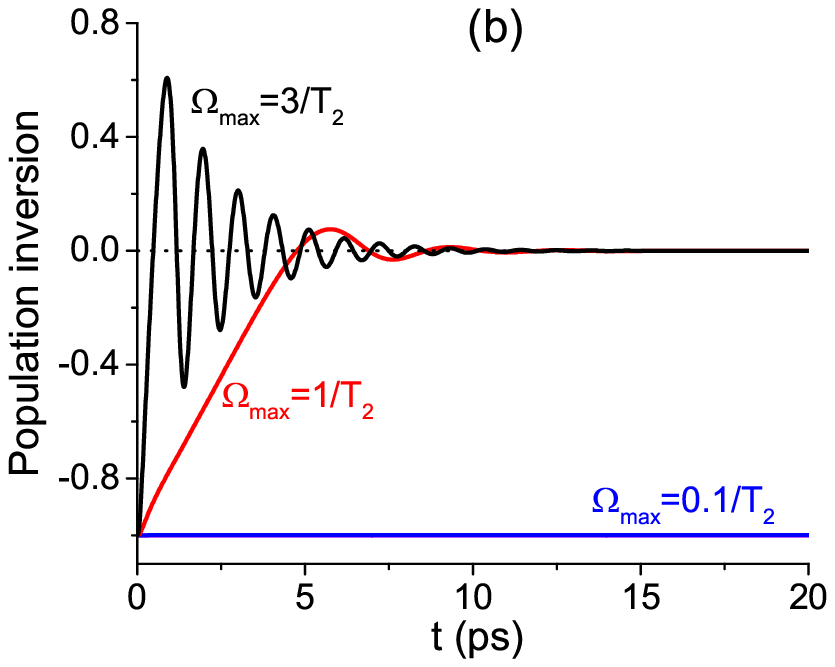}
\caption{\label{fig2} (a) Transmitted light intensity and (b) inversion dynamics at the entrance of the medium for different values of Rabi frequency. Parameters: $T_1=1$ ns, $T_2=1$ ps, $\omega_L=3.7 \cdot 10^{11}$ s$^{-1}$, $L=\lambda=1$ $\mu$m, $n=3$.}
\end{figure}

Transmission of light through the resonant medium strongly depends on its intensity, which can be conveniently characterized with the maximal Rabi frequency, $\Omega_{\textrm{max}}=\mu A/\hbar$, where $A$ is the wave amplitude. Figure \ref{fig2}(a) shows transmission dynamics for the continuous wave instantaneously switched on at $t=0$. One can see that the low-intensity radiation ($\Omega_{\textrm{max}} T_2 \ll 1$) is weakly transmitted through the layer of thickness $L=\lambda=1$ $\mu$m and is mostly absorbed inside the medium. On the contrast, the high-intensity radiation ($\Omega_{\textrm{max}} T_2 \sim 1$) is almost perfectly transmitted as it would be for the passive medium without absorbing particles at all. These two situations correspond to the steady-state inversions [Fig. \ref{fig2}(b)] equal to $w_{st} \approx -1$ (most two-level emitters remain in the ground state) and $w_{st} \approx 0$ (the equal number of emitters in the ground and excited states) and, therefore, will be referred to as ``absorbing medium'' and ``saturated medium'', respectively. Moreover, for high enough intensity ($\Omega_{\textrm{max}} T_2 > 1$), one can clearly observe the oscillations of transmission and the corresponding Rabi oscillations of population inversion. The inversion is given in Fig. \ref{fig2}(b) at the entrance of the medium, but it has essentially the same behavior at other distances due to small thickness of the layer. As an estimate, the condition $\Omega_{\textrm{max}} T_2 = 1$ requires the electric-field strength to be $E = \hbar / \mu T_2$, which for $T_2=1$ ps and $\mu = 30$ D (semiconductor quantum dots) gives $E = 10^6$ V/m corresponding to the intensity $\sim 5$ GW/m$^2$. This value can be further decreased for larger dipole moments and slower relaxation.

\section{Virtual gain condition}

In order to study the possibility of virtual gain, we consider the incident wave having constant intensity from $t=0$ to $t_{max}$ and exponentially decaying after that,
\begin{eqnarray}
E \sim A \left[ \theta(t_{max}-t) + e^{-(t-t_{max})/\tau} \theta(t-t_{max}) \right], \label{ampl}
\end{eqnarray}
where $\theta(t)$ is the Heaviside step function, $\tau$ is the decay time. The parameters of the system are chosen close to the pole, position of which in the complex-frequency plane can be evaluated analytically for a layer of thickness $L$ as follows
\begin{eqnarray}
\frac{\omega_p}{c} L = \frac{1}{n} \left[ \pi l +i \textrm{ln} \frac{n-1}{n+1} \right], \label{pole}
\end{eqnarray}
where $l$ is the integer number. The frequency of the wave \eqref{ampl} is tuned to coincide with the real part of pole frequency, $\omega=\omega'_p=2 \pi c/\lambda_p$. The decay time is, in turn, limited by the imaginary part of pole frequency, $\tau \lesssim \tau_p=1/\omega''_p$. One can see that for real $n$, $\lambda_p$ can be kept constant by simply changing simultaneously the thickness $L$ and the pole number $l$; further, we take $\lambda_p=1$ $\mu$m for $L=1$ $\mu$m, $l=6$, and $n=3$. On the contrary, $\omega''_p$ can be varied by changing $L$, so that for the above parameters we have $\tau_p \approx 0.014$ ps. This rather short time can be problematic to realize in experiment. However, it can be increased by taking thicker layer as shown further. As a proof of principle, we consider the thin layer, because it is easier and faster to calculate without significant influence of radiation inhomogeneity which can smear the effect in thick layers.

To give a simple explanation of the virtual gain in resonant media, it is convenient to use the simplified Bloch equations under RWA by adopting $s=0$ in Eqs. (\ref{polardl}) and (\ref{inversdl}). We also assume the exact resonance, so that $\omega=\omega_0$. First, we consider the stationary limit, when the radiation amplitude $\Omega$ is constant, whereas the polarization amplitude and population inversion have reached the steady state, $dp/dt=0$ and $dw/dt=0$. The steady-state values are as follows,
\begin{eqnarray}
p_{st}&=&\frac{i}{2} \Omega T_2 w_{st}, \label{polarstat} \\
w_{st}&=&-\frac{1}{1+|\Omega|^2 T_1 T_2}. \label{inverstat}
\end{eqnarray}
One can readily see that for low incident intensities ($|\Omega|^2 \ll (T_1 T_2)^{-1}$), the medium remains almost entirely unexcited, $w_{st} \approx -1$. On the contrary, for high intensity ($|\Omega|^2 \gg (T_1 T_2)^{-1}$), it is saturated, $w_{st} \approx 0$.

\begin{figure*}[t!]
\includegraphics[scale=0.95, clip=]{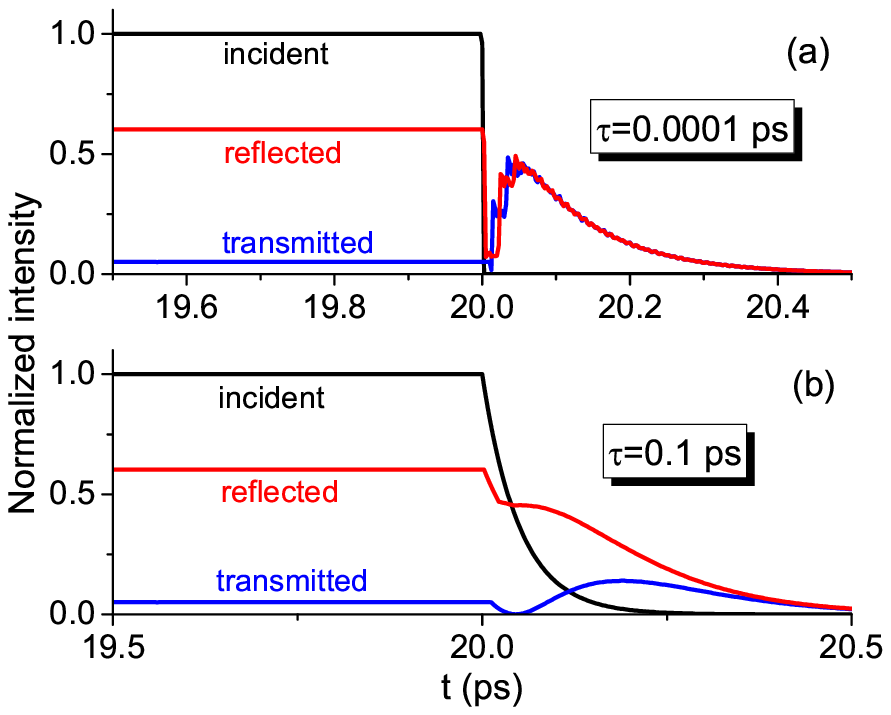}
\includegraphics[scale=0.95, clip=]{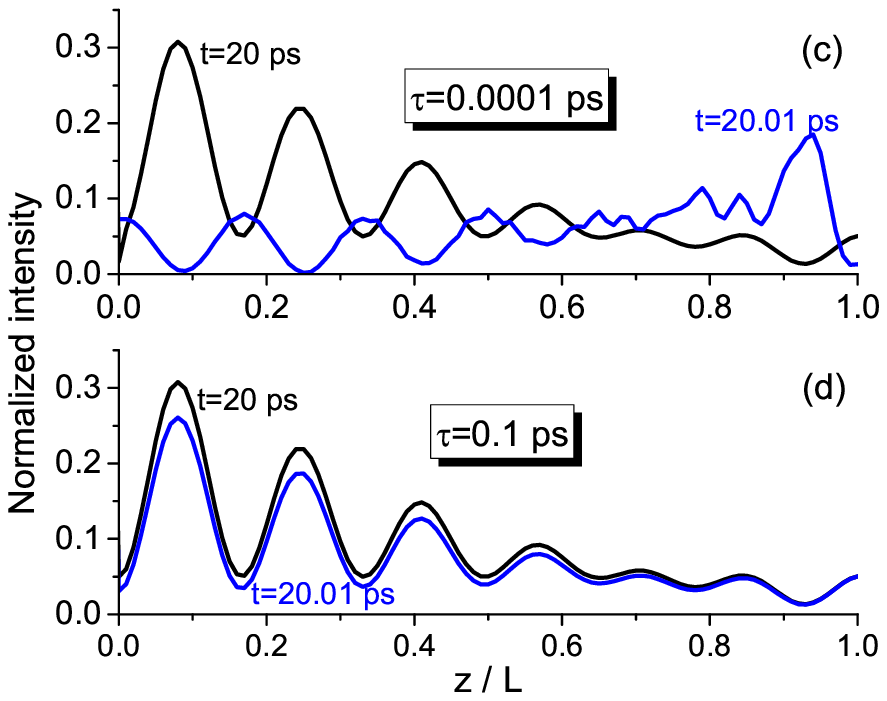}
\includegraphics[scale=0.95, clip=]{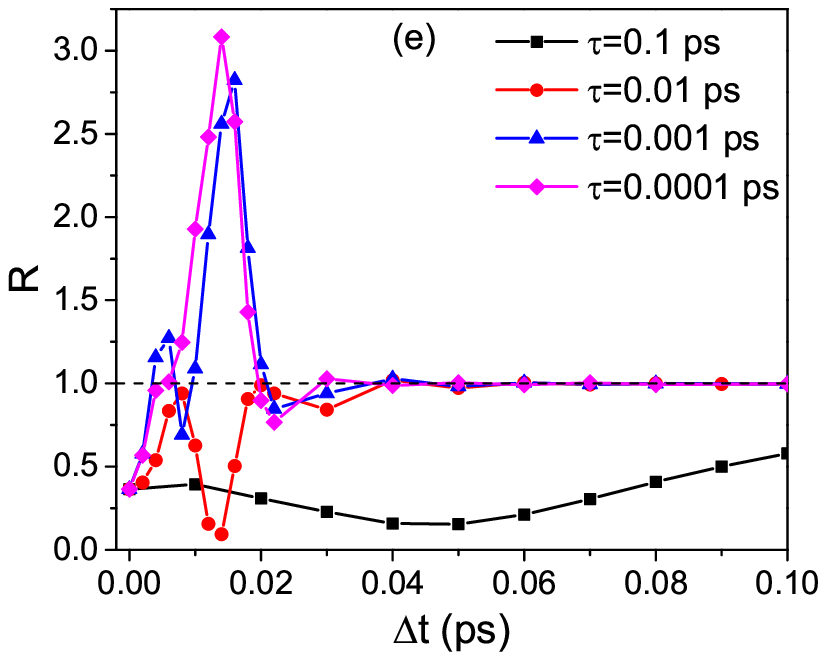}
\includegraphics[scale=0.95, clip=]{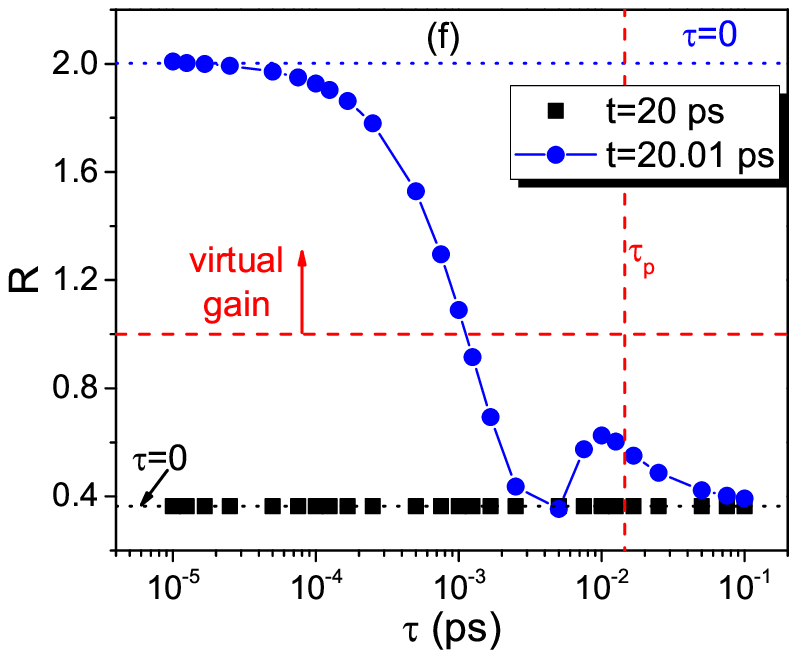}
\caption{\label{fig3} Virtual gain in the absorbing medium under low-intensity radiation, $\Omega_{\textrm{max}}=0.01/T_2$. (a) and (b) Light intensity profiles for different values of decay time $\tau$. (c) and (d) Radiation distributions at the switching-off moment ($t_{max}=20$ ps) and a short time later ($t_{max}+\Delta t=20.01$ ps). (e) The ratio $R$ as a function of $\Delta t$ for different $\tau$. (f) The ratio $R$ at different time instants as a function of the decay time $\tau$.}
\end{figure*}

If the radiation is switched off after the medium has reached steady state and decays exponentially ($\Omega=\Omega_0 e^{-t/\tau}$), we can assume that the polarization follows the decay of radiation as $p=p_{st} e^{-t/\tau}$. The population inversion, on the contrary, undergoes rapid decay due to the first term ($\sim e^{-2t/\tau}$) in the right-hand side of Eq. (\ref{inversdl}) and then changes slowly. So, we can assume that the inversion takes on a constant effective value, which can be found after substituting the exponential expressions for $\Omega$ and $p$ in Eq. (\ref{polardl}). Thus, we have
\begin{eqnarray}
w_{eff}=\left( 1-\frac{T_2}{\tau} \right) w_{st}. \label{invereff}
\end{eqnarray}
Obviously, $w_{eff}$ will have the sign opposite to the steady-state value $w_{st}$, if the decay time is short enough, $\tau<T_2$. This effect is especially pronounced in the absorbing medium regime, when $w_{st} \approx -1$ and $w_{eff}>0$, i.e., the medium acts effectively as if it would have gain. In the saturated medium regime, when $w_{st} \approx 0$, this effect should be negligible. Although this simplified explanation allows us to illustrate the possibility of virtual gain in the resonantly absorbing media, it cannot take into account the quantitative features due to the radiation propagation in the medium of finite length. 

Thus, we have several relations between the parameters of medium and radiation, which should be justified to observe virtual gain. Equation \eqref{pole} governs the position of pole and the corresponding limiting decay time $\tau_p$ as a function of thickness $L$ and refractive index $n$. Equation \eqref{invereff} requires $\tau<T_2$, so that one can expect virtual gain for $\tau \lesssim \tau_p <T_2$ in the absorbing medium regime governed by the condition $\Omega_{\textrm{max}} T_2 \ll 1$. One can easily assess the possible material parameters from these relations. Further, we justify this general reasoning with full numerical simulations of the Maxwell-Bloch equations in both the absorbing medium and saturated medium regimes.

\section{Absorbing medium regime}

We start with the regime of absorbing medium, when radiation is weak enough to saturate the medium. Figure \ref{fig3} shows the results of calculations for $\Omega_{\textrm{max}}=0.01/T_2$ and the incident wave given by Eq. \eqref{ampl} with the switching-off moment $t_{max}=20$ ps, i.e., the signal decays after the steady-state transmission and reflection are established. Behavior of radiation strongly depends on the decay time $\tau$: for slow decay with $\tau=0.1$ ps, transmitted and reflected intensities smoothly follow the profile of incident radiation [Fig. \ref{fig3}(b)], whereas for abruptly switched-off radiation with $\tau=10^{-4}$ ps, transmission and reflection demonstrate sharp splash in intensity [Fig. \ref{fig3}(a)]. The corresponding distributions of radiation along the medium just after the switching-off moment (at $t=t_{max}+\Delta t=20.01$ ps) are fundamentally different as compared in Figs. \ref{fig3}(c) and \ref{fig3}(d). For $\tau=0.1$ ps, the pattern of attenuating intensity characteristic for absorbing medium is kept after switching off the radiation. On the contrary, for $\tau=10^{-4}$ ps, the reversal of the pattern is seen with attenuation replaced by gradual growth of intensity despite the absorption in the material. This is exactly what we mean by the virtual gain in accordance with Ref. \cite{Li2020}.

We should emphasize that the virtual gain does not change the fact that the total amount of energy decreases with time as one would expect for a passive (absorbing) system \cite{Gu2022}. Moreover, the virtual gain is a transient phenomenon, i.e., the reversed pattern of intensity growing along the medium length exists only for a finite interval of time after radiation was switched off. Therefore, we need an integral value to illustrate the efficiency of virtual gain depending on decay time $\tau$ and how it changes with time. To this end, we calculate the ratio of average intensities in the second and the first halves of the layer,
\begin{eqnarray}
R(t) = \int_{L/2}^L I(t,z) dz / \int_0^{L/2} I(t,z) dz. \label{FOM}
\end{eqnarray}
Obviously, virtual gain corresponds to $R>1$, although this condition is not sufficient to make unequivocal conclusion on the regime of light propagation as will be clear further. Figure \ref{fig3}(e) shows how the ratio $R$ changes as a function of instant after the switching-off moment $\Delta t$ for different values of $\tau$. It is seen that for large $\tau$ (slow decay), $R$ stays lower than unity. Only for short enough $\tau$, the range of $\Delta t$ appears where $R(t)>1$ and virtual gain is possible. The transition to virtual gain agrees well with the condition of the pole position, $\tau \ll \tau_p$. Note that the effect is observed very soon after $t_{max}$ and $R$ rapidly (up to $0.1$ ps) comes to unity. The reasons are the fast redistribution of radiation leveling intensity along the medium and rapid exit of radiation from the medium. This uniformly distributed intensity grows in time as long as transmitted and reflected intensities increase according to Fig. \ref{fig3}(a). The ratio loses its meaning at very large $\Delta t$, because almost no radiation remains stored in the medium.

In Fig. \ref{fig3}(f), we show the dependence of $R$ on the decay time $\tau$ at the specific instant of time, $t_{max}+\Delta t=20.01$ ps. One can see that the ratio becomes larger than unity, when $\tau$ is much less than $\tau_p$. For $\tau=10^{-4}$ ps, it almost reaches the value obtained for instantaneous switching-off ($\tau=0$). Compare with the value of $R$ at $t_{max}=20$ ps which is the same for every $\tau$. Note that the value of $\tau_p$ is somewhat overestimated, because it is obtained for the real refractive index $n$, i.e., not taking into account medium absorption. Nevertheless, this simplified approach is good enough to give us the qualitative estimate of virtual gain possibility at short decay times.

\begin{figure}[t!]
\includegraphics[scale=0.95, clip=]{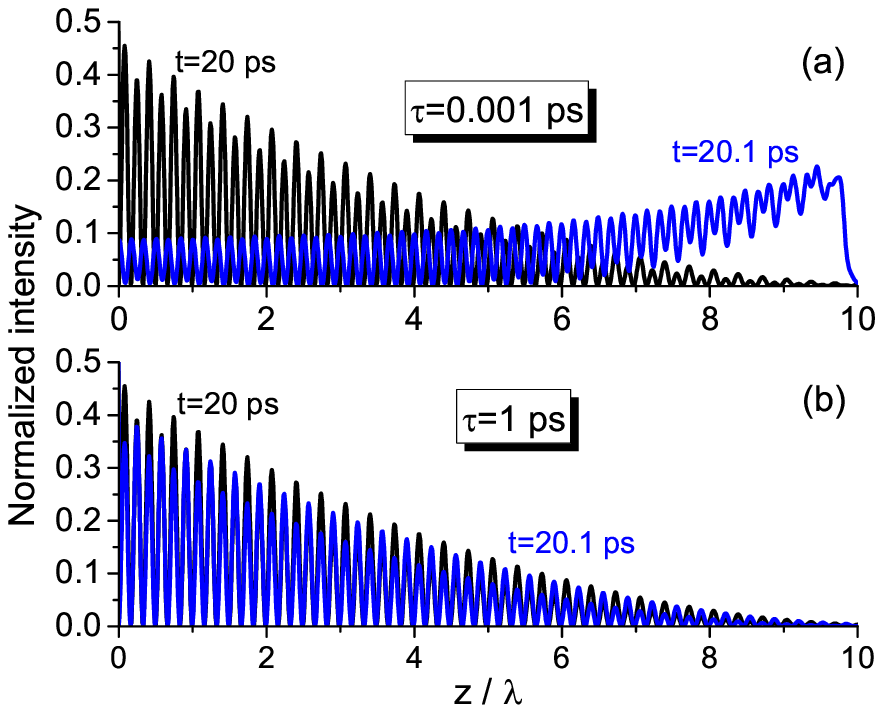}
\includegraphics[scale=0.95, clip=]{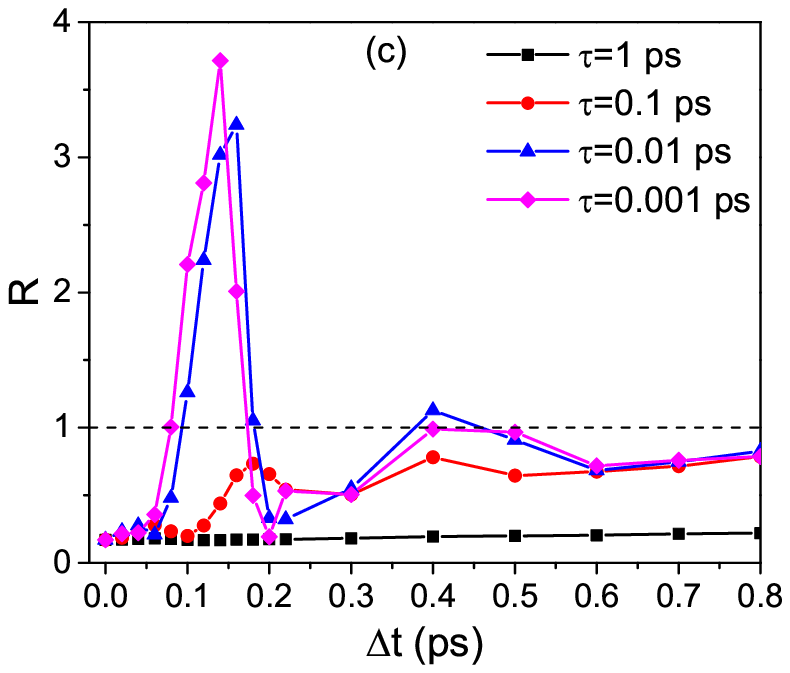}
\includegraphics[scale=0.95, clip=]{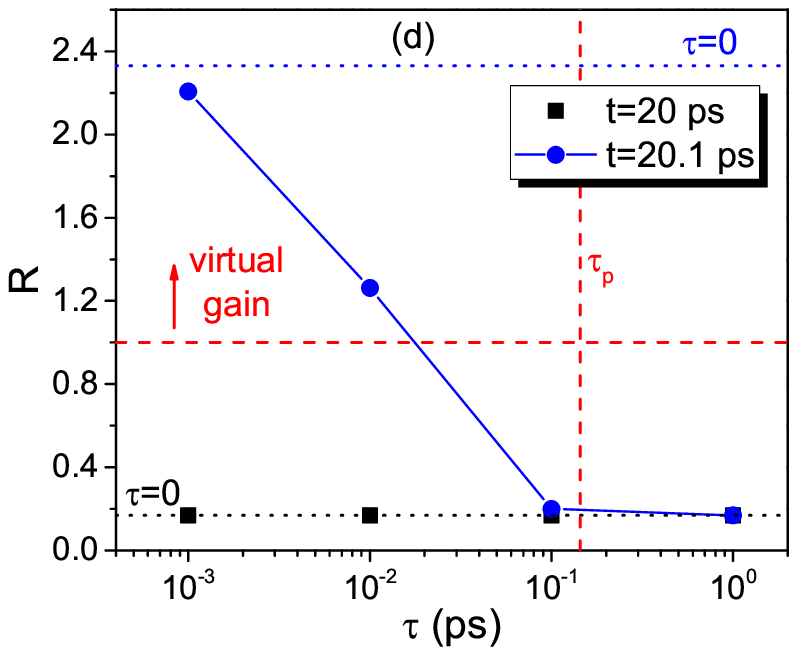}
\caption{\label{fig4} Virtual gain in the thick layer ($L=10 \lambda=10$ $\mu$m) of absorbing medium under low-intensity radiation, $\Omega_{\textrm{max}}=0.01/T_2$. (a) and (b) Radiation distributions at the switching-off moment ($t_{max}=20$ ps) and a short time later ($t_{max}+\Delta t=20.1$ ps). (c) The ratio $R$ as a function of $\Delta t$ for different decay times $\tau$. (d) The ratio $R$ at different time instants as a function of $\tau$.}
\end{figure}

As mentioned above, the value of $\tau_p$ grows with the medium thickness $L$, so that virtual gain can be reached at much higher decay times for a thick layer and may be easier observed in experiment. To give an example, we increase the thickness by $10$ times, up to $L=10 \lambda=10$ $\mu$m. As shown in Fig. \ref{fig4}(a), for fast enough decay ($\tau=10^{-3}$ ps), the intensity distribution soon after switching-off moment clearly demonstrates the transition from absorption to virtual gain. No such change is observed for slow decay [$\tau=1$ ps, Fig. \ref{fig4}(b)]. Corresponding behavior of the ratio $R$ [Figs. \ref{fig4}(c) and \ref{fig4}(d)] also corroborates the validity of observations made for the thin layer. All the time values such as $\Delta t$ corresponding to the $R$ peak in Fig. \ref{fig4}(c) increase by an order of magnitude. The same can be said about the decay time, so that $\tau=0.01$ ps $\ll \tau_p=0.14$ ps is enough for virtual gain. One can expect that further increase of the layer thickness can relax the requirements on the decay rapidity even stronger.

\section{Saturated medium regime}

\begin{figure}[t!]
\includegraphics[scale=0.95, clip=]{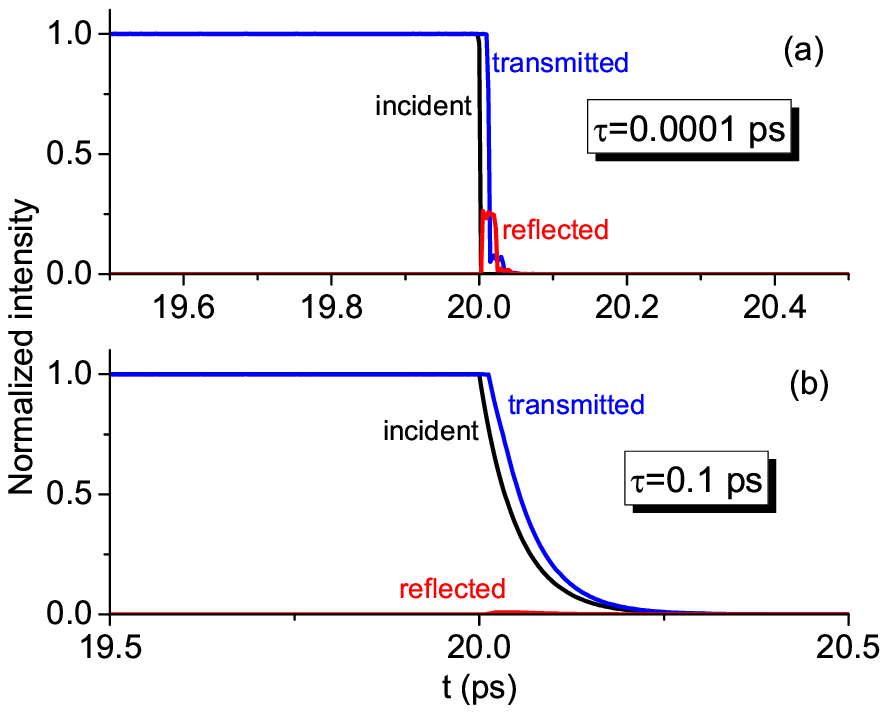}
\includegraphics[scale=0.95, clip=]{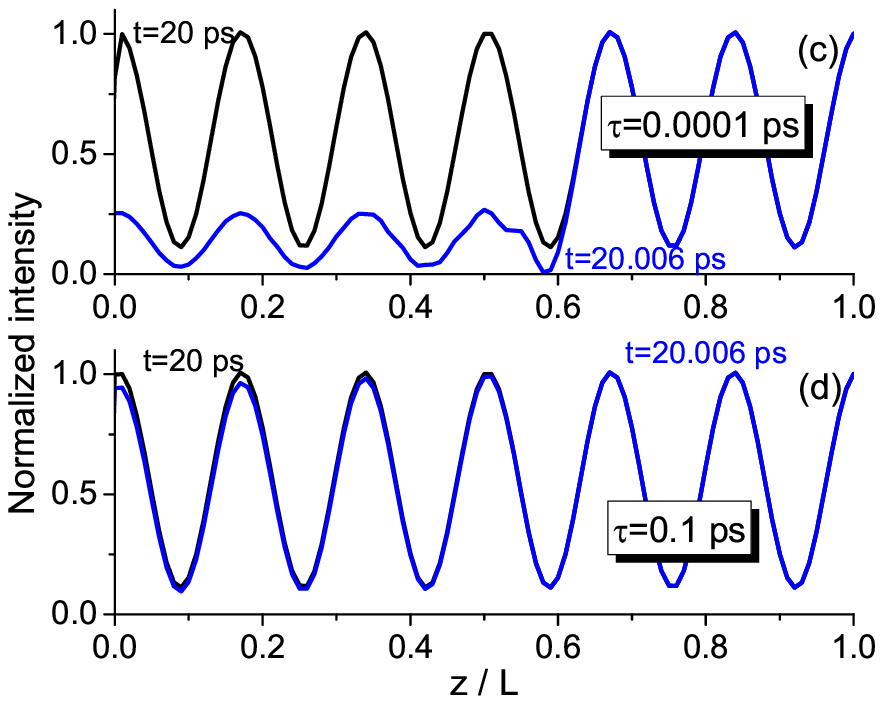}
\caption{\label{fig5} The case of saturated medium under high-intensity radiation, $\Omega_{\textrm{max}}=3/T_2$. (a) and (b) Light intensity profiles for different values of decay time. (c) and (d) Radiation distributions at the switching-off moment ($t_{max}=20$ ps) and a short time later ($t_{max}+\Delta t=20.006$ ps).}
\end{figure}

\begin{figure}[t!]
\includegraphics[scale=0.95, clip=]{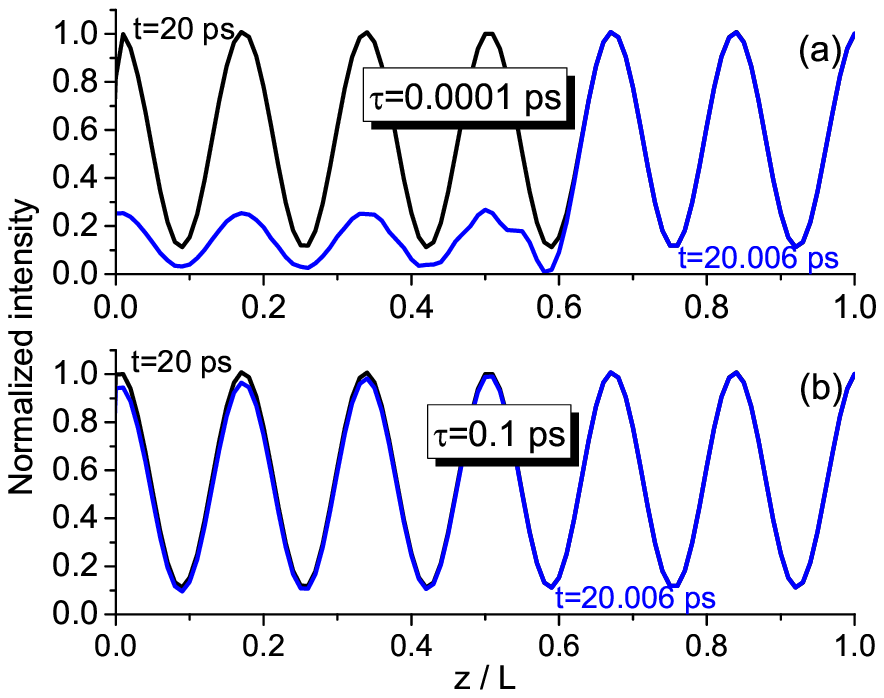}
\includegraphics[scale=0.95, clip=]{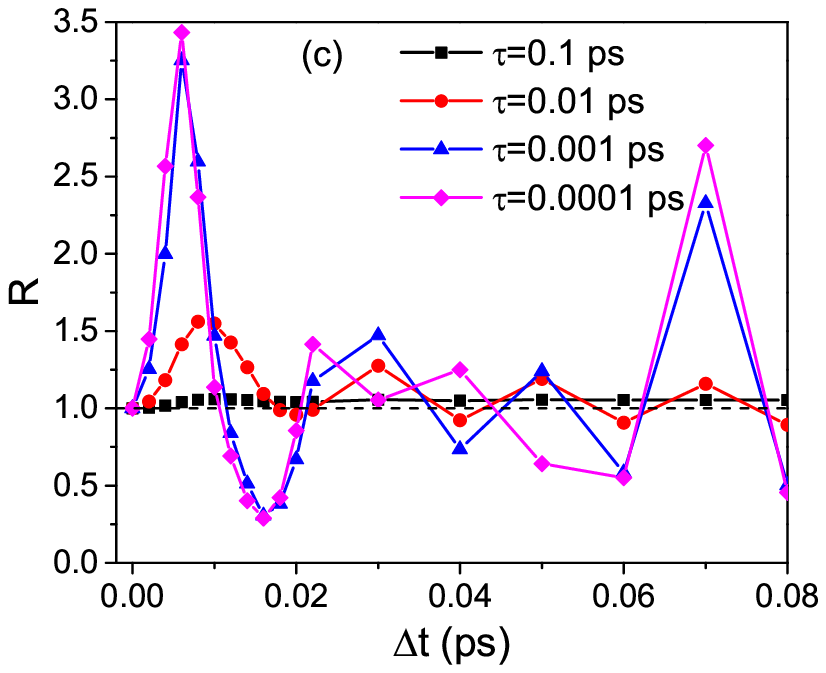}
\caption{\label{fig6} The case of passive medium under low-intensity radiation, $\Omega_{\textrm{max}}=0.01/T_2$. (a) and (b) Radiation distributions at the switching-off moment ($t_{max}=20$ ps) and a short time later ($t_{max}+\Delta t=20.006$ ps). (c) The ratio $R$ as a function of $\Delta t$ for different $\tau$.}
\end{figure}

\begin{figure*}[t!]
\includegraphics[scale=0.95, clip=]{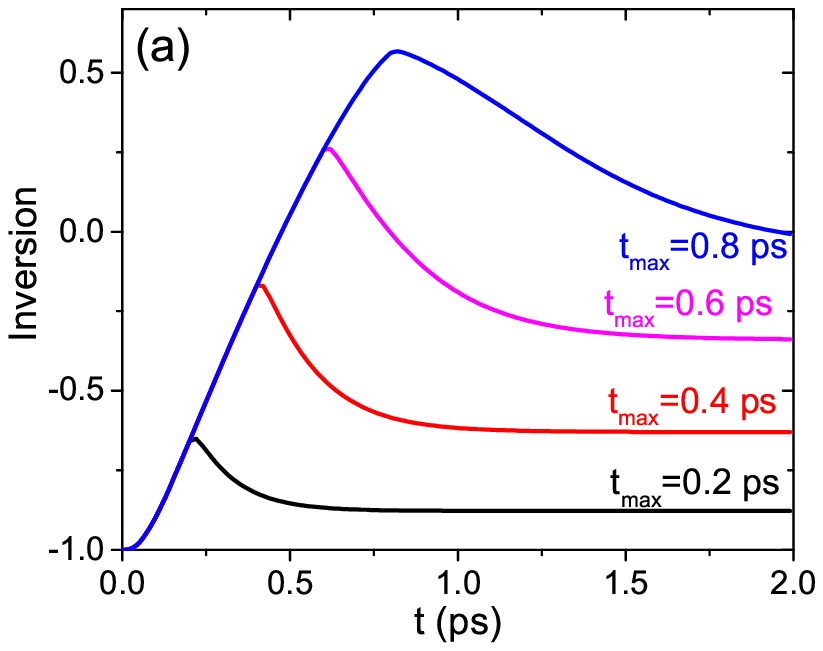}
\includegraphics[scale=0.95, clip=]{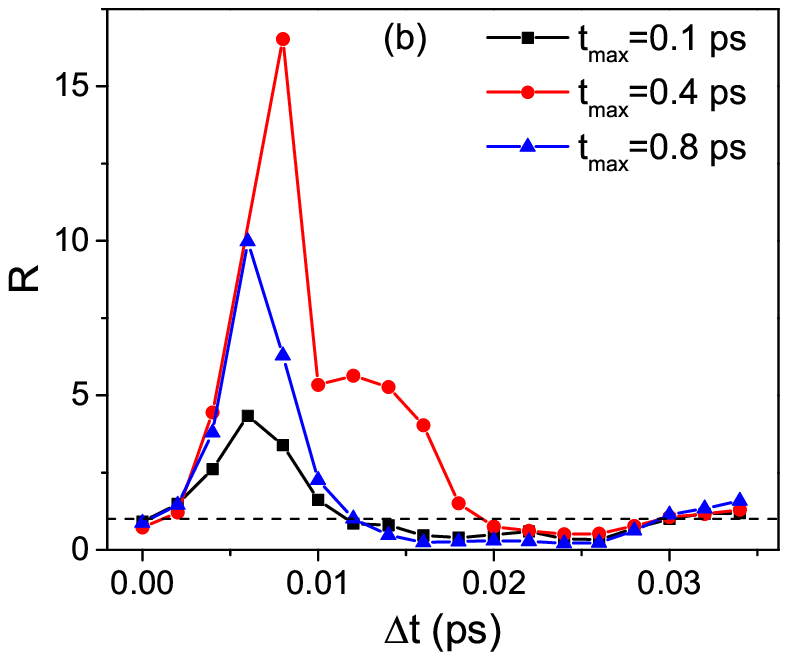}
\includegraphics[scale=0.95, clip=]{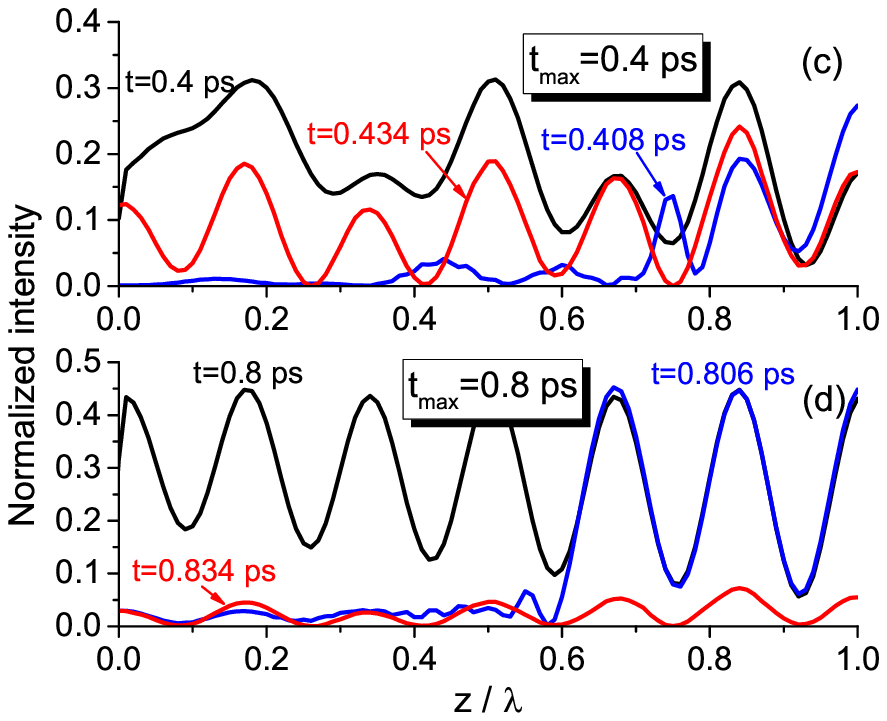}
\includegraphics[scale=0.95, clip=]{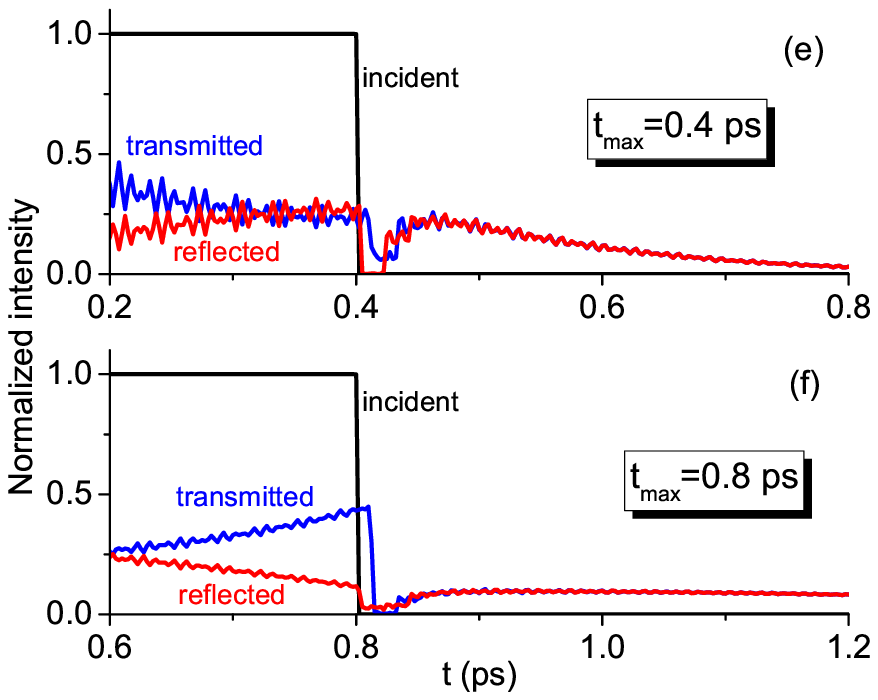}
\includegraphics[scale=0.95, clip=]{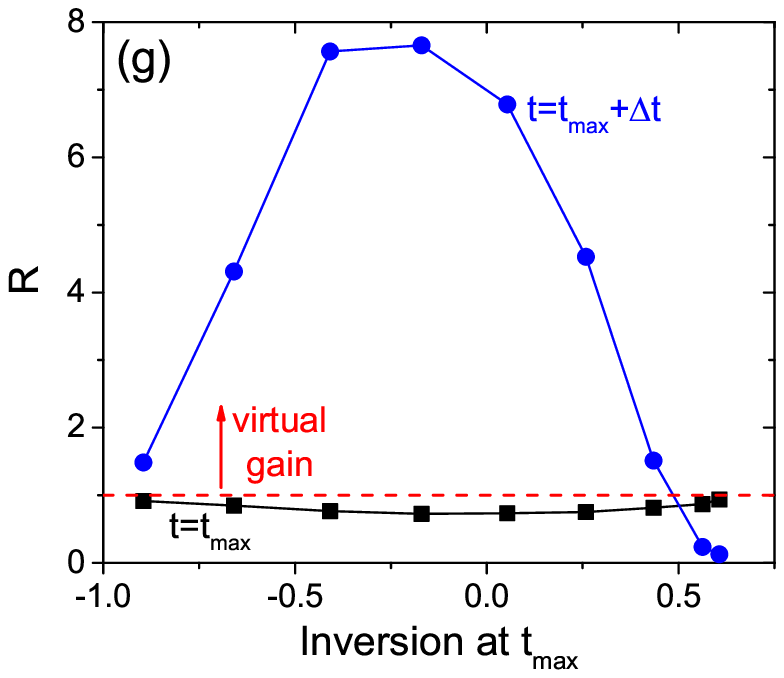}
\caption{\label{fig7} Tunable virtual gain under high-intensity radiation, $\Omega_{\textrm{max}}=3/T_2$. (a) Dynamics of population inversion for different switching-off moments $t_{max}$. (b) The ratio $R$ as a function of $\Delta t$ for different $t_{max}$. (c) and (d) Radiation distributions at the switching-off moment $t_{max}$ and a short time later (at $t=t_{max}+\Delta t$ marked in the plot). (e) and (f) Light intensity profiles for different switching-off moments $t_{max}$. (g) The ratio $R$ as a function of inversion at $t_{max}$ calculated at the instant $t=1.02 t_{max}$. The decay time is $\tau=10^{-4}$ ps.}
\end{figure*}

Let us now consider the case of radiation powerful enough to saturate the medium. In particular, we take the amplitude $\Omega_{\textrm{max}}=3/T_2$. Radiation is switched off at $t_{max}=20$ ps, when the steady state is already reached, i.e., the population inversion is close to zero and transmission is close to unity (see Fig. \ref{fig2}). The key difference in this case is the absence of absorption in the medium. As a result, light at the switching-off moment is evenly distributed along the layer, without any attenuation (Fig. \ref{fig5}). If radiation is decaying slowly, this uniform distribution is saved [Fig. \ref{fig5}(d)] as well as unity transmission, i.e., radiation gradually leaves the medium only in the forward direction. If radiation is decaying rapidly, the distribution takes the form of a sharp step in intensity moving along the medium in the forward direction [Fig. \ref{fig5}(c)]. The sharp steps are observed also in the transmitted and reflected intensity profiles as shown in Fig. \ref{fig5}(a). The profiles for transmitted radiation generally follow the incident intensity. The difference between slow and fast decays is that in the latter case, the profile forms a step due to very sharp disappearance of impinging radiation [Fig. \ref{fig5}(a)]. The corresponding step is observed also in reflection, unlike almost total absence of reflected signal in the case of slow decay [Fig. \ref{fig5}(b)]. The stepped distribution makes a fundamental difference between the saturated medium regime and the case of absorbing medium with its monotonously growing intensity [compare Figs. \ref{fig5}(c) and \ref{fig3}(c)] and allows us to conclude that the saturated medium does not support virtual gain.

In fact, the saturated medium regime is equivalent to the case of purely passive, lossless medium, which contains no absorbing particles as illustrated by the characteristic step distribution [Fig. \ref{fig6}(a), compare with Fig. \ref{fig5}(c)]. In Fig. \ref{fig6}(c), we plot the ratio $R$ at different time instants showing the sharp peak soon after the switching-off moment. This peak is clearly associated with the step in the intensity profile [Fig. \ref{fig5}(a)] and is due to radiation rapidly leaving the medium. The strong fluctuations at $\Delta t >0.2$ ps illustrate the meaninglessness of $R$ at later times because there is almost no radiation inside the layer. Note also that $R$ remains around unity for slow decay ($\tau=0.1$ ps) that corresponds to the even distribution of radiation along the medium [Fig. \ref{fig6}(b)]. In the case of saturated medium, the ratio demonstrates the same behavior as in Fig. \ref{fig6}(c).

In addition, in Appendix \ref{nopole}, we discuss the system, which does not satisfy the condition for pole \eqref{pole}, and show that it can be considered as an intermediary case having features of both absorbing and saturated medium regimes.

\section{Nonstationary regime}

In the above consideration, we have assumed that radiation is switched off only after the steady state of both transmission and population inversion is established. Here, we turn to the nonstationary case, when decay starts before the steady state is reached. This opens a room for controlling the virtual response by choosing the proper value of the switching-off moment $t_{max}$. To demonstrate this possibility, it is instructive to take the high-intensity incident wave able to strongly change the medium state in a short period of time. We concentrate at the rising slope of the first Rabi oscillation seen Fig. \ref{fig2}. In Fig. \ref{fig7}(a), we show the population-inversion dynamics for $\Omega_{\textrm{max}}=3/T_2$ and different $t_{max}$, but for the same decay time $\tau=10^{-4}$ ps. Changing $t_{max}$ in the range of just $1$ ps is enough to get different values of inversion $w(t_{max})$ at the switching-off moment and, as a result, different behavior of radiation inside the medium.

The dynamics of ratio $R$ as a function of time after switching-off radiation demonstrate the sharp peaks with the height depending on $t_{max}$ [Fig. \ref{fig7}(b)]. It is maximal for $t_{max}$ corresponding to the population inversion close to zero and diminishes for large absolute values of $w(t_{max})$. However, the behavior hiding behind these peaks seems to be quite different as evidenced by the comparison of radiation distributions for $t_{max}=0.4$ ps [Fig. \ref{fig7}(c)] and $t_{max}=0.8$ ps [Fig. \ref{fig7}(d)], respectively. In the first case, the nonperiodic (because of nonstationarity) distribution at $t=t_{max}=0.4$ ps is replaced by growing intensity characteristic for virtual gain at $t=0.408$ ps. In the second case, the initial distribution at $t=t_{max}=0.8$ ps is more periodic (the system seems to be closer to the steady state), whereas the distribution at $t=0.806$ ps (maximal $R$) is step-like as in the saturated-medium regime. At later times [$\Delta t=0.034$ ps in Figs. \ref{fig7}(c) and \ref{fig7}(d)], the distributions are more uniform and lower in intensity. The corresponding intensity profiles are shown in Figs. \ref{fig7}(e) and \ref{fig7}(f) for different switching-off moments $t_{max}$ at the same decay time $\tau=10^{-4}$ ps. For $t_{max}=0.4$ ps, the population inversion is negative [see Fig. \ref{fig7}(a)], so that transmission is decreasing as expected for absorbing medium. For $t_{max}=0.8$ ps, the population inversion gets positive and leads to increasing transmission. As a result, in the latter case, the switching-off moment is followed by the clear step-like feature and subsequent very slow intensity decay. These peculiarities of intensity profiles are linked to the differences between the distributions shown in Fig. \ref{fig7}(c) and \ref{fig7}(d).

The dependence of $R$ calculated at the instant $t=1.02 t_{max}$ on the population inversion at the switching-off moment $t_{max}$ [Fig. \ref{fig7}(g)] clearly demonstrates that virtual gain is most effective for $w(t_{max}) \lesssim 0$. For positive $w(t_{max})$, the ratio $R$ rapidly decreases and the response changes its character from virtual gain to saturated-medium-like regime as discussed above.

Thus, for $w(t_{max})<0$, the medium can be considered as absorbing and we obtain virtual gain. On the contrary, for $w(t_{max})>0$, the situation is analogous to the case of saturated medium and the peak in $R$ is not connected to the virtual gain, but is due to the step-like distribution. Regulating the value of $t_{max}$ allows one to control the regime of light-matter interaction and switch virtual gain on and off. This idea can be used as the tunability scheme allowing to obtain response corresponding to different effective gains controlled simply by the switching-off moment $t_{max}$ of decaying radiation.

\section{Conclusion}

Our results corroborate the possibility to harness scattering anomalies (such as poles) to obtain unusual response under optical irradiation strongly varying in time. We demonstrate that resonantly absorbing media illuminated with an exponentially decaying light can be used to realize virtual gain with the efficiency tuned dynamically using intensity-driven population inversion. Simultaneously, one can use this scheme to control the medium state changing its variables (population inversion and polarization) in a desired way. There is a certain paradox that virtual gain is most pronounced in the presence of absorption and not in the passive or saturated medium. It would be intriguing to further generalize our approach to the initially excited media or combination of several layers of different resonant media.

\acknowledgements{The work has been supported by the Belarusian Republican Foundation for Fundamental Research (Project No. F22TURC-001). The author is grateful to Viktoryia Kouhar for help with preparation of the graphical materials.}

\appendix

\section{\label{RWA}Details on the Maxwell-Bloch equations and applicability of the RWA}

\begin{figure}[t!]
\includegraphics[scale=0.95, clip=]{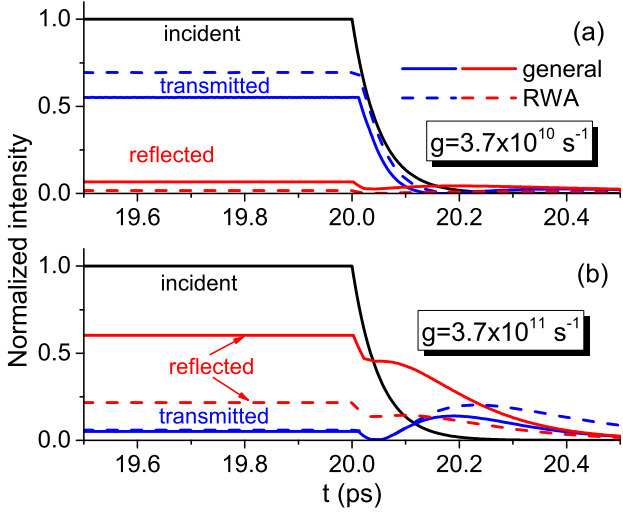}
\caption{\label{fig8} Testing RWA applicability: Light intensity profiles at $\tau=0.1$ ps for different values of the Lorentz frequency: (a) $\omega_L=3.7 \cdot 10^{10}$ s$^{-1}$ and (b) $\omega_L=3.7 \cdot 10^{11}$ s$^{-1}$.}
\end{figure}

\begin{figure}[t!]
\includegraphics[scale=0.95, clip=]{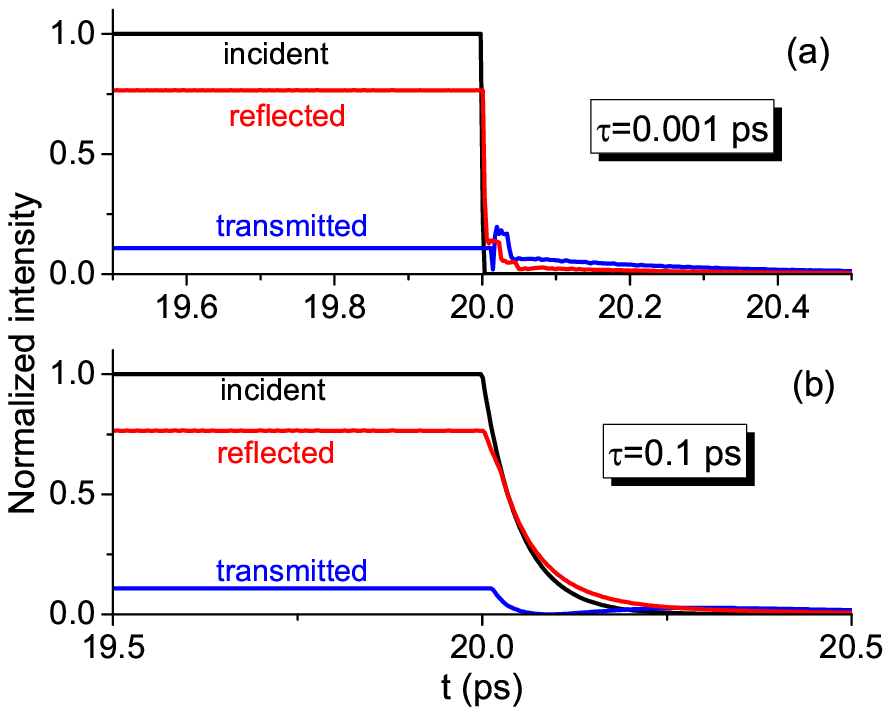}
\includegraphics[scale=0.95, clip=]{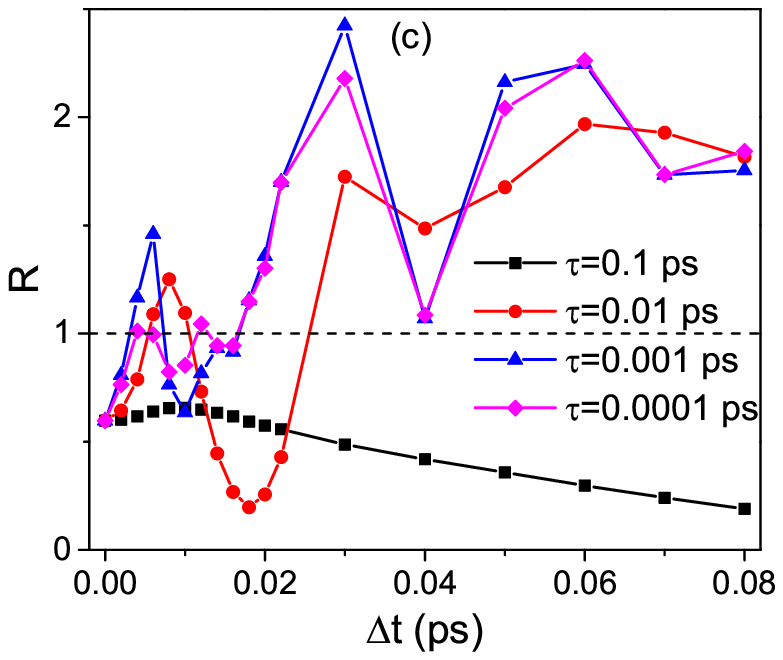}
\includegraphics[scale=0.95, clip=]{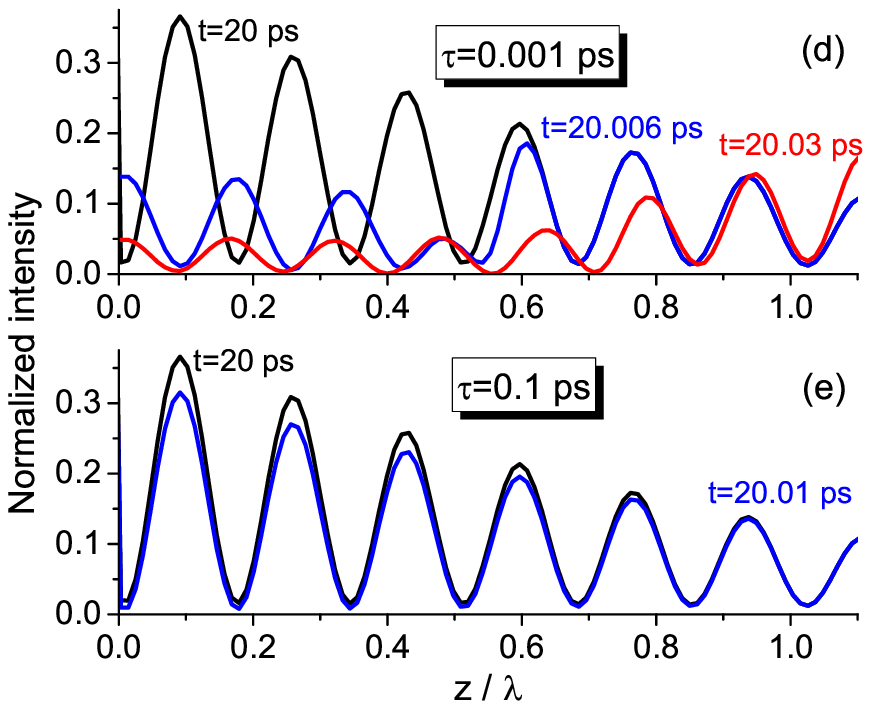}
\caption{\label{fig9} Virtual gain in the layer not satisfying the pole condition ($L=1.1 \lambda=1.1$ $\mu$m) under low-intensity radiation, $\Omega_{\textrm{max}}=0.01/T_2$. (a) and (b) Light intensity profiles for different values of decay time $\tau$. (c) The ratio $R$ as a function of $\Delta t$ for different $\tau$. (d) and (e) Radiation distributions for different $\tau$ at the switching-off moment ($t_{max}=20$ ps) and a short time later ($t=t_{max}+\Delta t$).}
\end{figure}

The applicability of the rotating-wave approximation (RWA) is convenient to study with the approach described in Ref. \cite{Novitsky2012}. According to this approach, the Maxwell-Bloch equations can be represented in the dimensionless form as follows,
\begin{eqnarray}
\frac{\partial^2 \Omega'}{\partial \xi^2}&-& \frac{\partial^2
\Omega'}{\partial \theta^2}-2 i \frac{\partial \Omega'}{\partial \xi}-2
i \frac{\partial \Omega'}{\partial
\theta} + (n^2-1) \Omega' \nonumber \\
&&=6 \epsilon \left(\frac{\partial^2 p}{\partial \theta^2}+2 i
\frac{\partial p}{\partial \theta}-p\right), \label{Maxdl}
\end{eqnarray}
\begin{eqnarray}
\frac{d p}{d \theta} &=& i \delta p + \frac{i}{2} (\Omega' + s \Omega'^*
e^{-2 i (\theta-\xi)}) w - \gamma'_2 p, \label{polardl}
\end{eqnarray}
\begin{eqnarray}
\frac{d w}{d \theta} &=& i (\Omega'^* p - \Omega' p^*) + i s
\left(\Omega' p
e^{2 i (\theta-\xi)} - \Omega'^* p^* e^{-2 i (\theta-\xi)} \right) \nonumber \\
&&- \gamma'_1 (w+1), \label{inversdl}
\end{eqnarray}
where we represented the electric field and atomic polarization as $E= \{ A \exp [i (\omega t - k z)] + \textrm{c.c.} \} /2$ and $\rho_{12}=p \exp [i (\omega t - k z)]$. Here $\omega$ is the carrier frequency of radiation, $k=\omega/c$ is the wavenumber, $\theta=\omega t$ and $\xi=kz$ are dimensionless
arguments, $\Omega'=(\mu/\hbar \omega) A$ is the normalized Rabi frequency, $\delta=\Delta \omega/\omega=(\omega_0-\omega)/\omega$ is the
frequency detuning, $\gamma'_{1,2}=1/T_{1,2}\omega$ are the
normalized relaxation rates, $\epsilon= g/\omega = 4 \pi
\mu^2 C/3 \hbar \omega$ is the strength of light-matter interaction (normalized Lorentz frequency), and $\textrm{c.c.}$ stands for complex conjugate. The numerical method to solve these equations is essentially the same as in Ref. \cite{Novitsky2009}.

The key factor governing account of RWA is the auxiliary parameter $s$: when it is dropped out ($s=0$), we have the usual RWA; otherwise, the rapidly rotating terms are taken into account. The problem of RWA applicability is very well studied for short pulses. It is justified for coherent pulses with duration much shorter than the relaxation times, but for very short, few-cycle pulses it is violated \cite{Novitsky2012}. For continuous or quasi-continuous radiation as in our case, the RWA violation can be regulated by the values of the relaxation times $T_{1,2}$ and light-matter-interaction strength, $g=4 \pi \mu^2 C/3 \hbar$. It is known that the RWA can be violated in the strong-coupling regime, when Rabi oscillations become possible \cite{Hughes1998}. An example of intensity profile calculations for different $g$ is shown in Fig. \ref{fig8}. One can see that decreasing $g$ makes the general and RWA cases closer to each other. Simultaneously, transmission grows and absorption decreases due to weaker light-matter interaction. Since we deal with a rather thin layer ($L=\lambda$), the large value of $g$ is justified to observe the pronounced effects. Nevertheless, the qualitative features of virtual gain are observed both in the general case and under the RWA.

\section{\label{nopole}Virtual gain in the system not satisfying the pole condition}

Here, we take the layer of slightly larger thickness as compared to previous consideration, so that one cannot find the integer $l$ to match the pole condition \eqref{pole}. In particular, we take $L=1.1 \lambda=1.1$ $\mu$m. The results of calculations are shown in Fig. \ref{fig9}. One can see that there is still strong dependence on the decay time $\tau$, although virtual gain can be observed in not so sharp and clear form as at the pole. For large $\tau$, transmission and reflection follow the incident intensity [Fig. \ref{fig9}(b)] and absorption is prevailing even after switching the signal off [Fig. \ref{fig9}(e)], so that the ratio $R$ remains lower than unity [Fig. \ref{fig9}(c), black line]. On the contrary, $R$ for short $\tau$ demonstrates several peaks at different time instants [see, e.g., blue line in Fig. \ref{fig9}(c)]. These peaks have different origin: the first one corresponds to the step-like distribution of radiation, whereas the second one more resembles virtual gain [compare blue and red curves in Fig. \ref{fig9}(d)]. The profile shown in Fig. \ref{fig9}(a) corroborates this interpretation containing both the sharp step-like splash of intensity and the subsequent smooth decay of transmitted signal. Thus, we can still have the virtual gain outside the pole, but not so effective as exactly at the pole.


\begin{thebibliography}{00}
\bibitem{GaponenkoBook} S. V. Gaponenko, Introduction to Nanophotonics (Cambridge University Press, Cambridge, 2010).

\bibitem{Krasnok2019} A. Krasnok, D. Baranov, H. Li, M.-A. Miri, F. Monticone, and A. Al\`{u}, Anomalies in light scattering, {Adv. Opt. Photon.} \textbf{11}, 892 (2019).

\bibitem{Chong2010} Y. D. Chong, L. Ge, H. Cao, and A. D. Stone, Coherent Perfect Absorbers: Time-Reversed Lasers, {\prl} \textbf{105}, 053901 (2010).

\bibitem{Longhi2010} S.~Longhi, $\mathcal{PT}$-symmetric laser absorber, {\pra} \textbf{82}, 031801(R) (2010).

\bibitem{Wong2016} Z.~J.~Wong, Y.-L.~Xu, J.~Kim, K.~O'Brien, Y.~Wang, L.~Feng, and X.~Zhang, Lasing and anti-lasing in a single cavity, {Nat. Photon.} \textbf{10}, 796 (2016).

\bibitem{Hsu2016} C. W. Hsu, B. Zhen, A. D. Stone, J. D. Joannopoulos, and M. Solja\v{c}i\'{c}, Bound states in the continuum, {Nat. Rev. Mater.} \textbf{1}, 16048 (2016).

\bibitem{Novitsky2022} D. V. Novitsky, A. Can\'os Valero, A. Krotov, T. Salgals,
A. S. Shalin, and A. V. Novitsky, CPA-lasing associated with the quasibound states in the continuum in asymmetric non-Hermitian structures, {ACS Photon.} \textbf{9}, 3035 (2022).

\bibitem{Miri2019} M.-A. Miri and A. Al\`{u}, Exceptional points in optics and photonics, {Science} \textbf{363}, eaar7709 (2019).

\bibitem{Baryshnikova2019} K. V. Baryshnikova, D. A. Smirnova, B. S. Luk'yanchuk, Yu. S. Kivshar, Optical Anapoles: Concepts and Applications, {Adv. Opt. Photon.} \textbf{7}, 1801350 (2019).

\bibitem{Tribelsky2006} M. I. Tribelsky and B. S. Luk’yanchuk, Anomalous Light Scattering by Small Particles, {\prl} \textbf{97}, 263902 (2006).

\bibitem{Ruan2010} Z. Ruan and S. Fan, Superscattering of Light from Subwavelength Nanostructures, {\prl} \textbf{105}, 013901 (2010).

\bibitem{Galiffi2022} E. Galiffi, R. Tirole, S. Yin, H. Li, S. Vezzoli, P. A. Huidobro, M. G. Silveirinha, R. Sapienza, A. Al\`{u}, and J. B. Pendry, Photonics of time-varying media, {Adv. Photon.} {\bf4}, 014002 (2022).

\bibitem{Xiao2014} Y. Xiao, D. N. Maywar, and G. P. Agrawal, Reflection and transmission of electromagnetic waves at a temporal boundary, {\ol} {\bf39}, 574 (2014).

\bibitem{Plansinis2015} B. W. Plansinis, W. R. Donaldson, and G. P. Agrawal, What is the Temporal Analog of Reflection and Refraction of Optical Beams?, {\prl} {\bf115}, 183901 (2015).

\bibitem{Ramaccia2020} D. Ramaccia, A. Toscano, and F. Bilotti, Light propagation through metamaterial temporal slabs: reflection, refraction, and special cases, {\ol} {\bf45}, 5836 (2020).

\bibitem{Biancalana2007} F. Biancalana, A. Amann, A. V. Uskov, and E. P. O’Reilly, Dynamics of light propagation in spatiotemporal dielectric structures, {\pre} {\bf75}, 046607 (2007).

\bibitem{Yin2022} S. Yin, E. Galiffi, and A. Al\`{u}, Floquet metamaterials, {eLight} {\bf2}, 8 (2022).

\bibitem{Mostafa2022} M. H. Mostafa, A. D\'{i}az-Rubio, M. S. Mirmoosa, and S. A. Tretyakov, Coherently Time-Varying Metasurfaces, {Phys. Rev. Appl.} {\bf17}, 064048 (2022).

\bibitem{Pacheco2021} V. Pacheco-Pe\~{n}a and N. Engheta, Temporal equivalent of the Brewster angle, {\prb} {\bf104}, 214308 (2021).

\bibitem{Sharabi2021} Y. Sharabi, E. Lustig, and M. Segev, Disordered Photonic Time Crystals, {\prl} {\bf126}, 163902 (2021).

\bibitem{Lustig2018} E. Lustig, Y. Sharabi, and M. Segev, Topological aspects of photonic time crystals, {Optica} {\bf5}, 1390 (2018).

\bibitem{Li2021} H. Li, S. Yin, E. Galiffi, and A. Al\`{u}, Temporal Parity-Time Symmetry for Extreme Energy Transformations, {\prl} {\bf127}, 153903 (2021).

\bibitem{Lasri2022} O. Lasri and L. Sirota, Temporal negative refraction, arXiv:2209.10647 (2022).

\bibitem{Morgenthaler1958} F. R. Morgenthaler, Velocity Modulation of Electromagnetic Waves, {IRE Trans. Microwave Theory Tech.} {\bf6}, 167 (1958).

\bibitem{Holberg1966} D. Holberg and K. Kunz, Parametric properties of fields in a slab of time-varying permittivity, {IEEE Trans. Ant. Prop.} {\bf14}, 183 (1966).

\bibitem{Lyubarov2022} M. Lyubarov, Y. Lumer, A. Dikopoltsev, E. Lustig, Y. Sharabi, and M. Segev, Amplified emission and lasing in photonic time crystals, {Science} {\bf377}, 425 (2022).

\bibitem{Sharabi2022} Y. Sharabi, A. Dikopoltsev, E. Lustig, Y. Lumer, and M. Segev, Spatiotemporal photonic crystals, {Optica} {\bf9}, 585 (2022).

\bibitem{Wang2022} X. Wang, M. S. Mirmoosa, V. S. Asadchy, C. Rockstuhl, S. Fan, and S. A. Tretyakov, Metasurface-Based Realization of Photonic Time Crystals, arXiv:2208.07231 (2022).

\bibitem{Hayran2022} Z. Hayran, J. B. Khurgin, and F. Monticone, $\hbar \omega$ versus $\hbar k$: dispersion and energy constraints on time-varying photonic materials and time crystals, {Opt. Mater. Express} {\bf12}, 3904 (2022).

\bibitem{Baranov2017} D. G. Baranov, A. Krasnok, and A. Al\`{u}, Coherent virtual absorption based on complex zero excitation for ideal light capturing, {Optica} {\bf4}, 1457 (2017).

\bibitem{Longhi2018} S.~Longhi, Coherent virtual absorption for discretized light, {\ol} \textbf{43}, 2122 (2018).

\bibitem{Zhong2020} Q. Zhong, L. Simonson, T. Kottos, and R. El-Ganainy, Coherent virtual absorption of light in microring resonators, {Phys. Rev. Res.} \textbf{2}, 013362 (2020).

\bibitem{Marini2020} A. Marini, D. Ramaccia, A. Toscano, and F. Bilotti, Metasurface-bounded open cavities supporting virtual absorption: free-space energy accumulation in lossless systems, {\ol} \textbf{45}, 3147 (2020).

\bibitem{Trainiti2019} G. Trainiti, Y. Ra’di, M. Ruzzene, and A. Al\`{u}, Coherent virtual absorption of elastodynamic waves, {Sci. Adv.} {\bf5}, eaaw3255 (2019).

\bibitem{Li2020} H. Li, A. Mekawy, A. Krasnok, and A. Al\`{u}, Virtual parity-time symmetry, {\prl} {\bf124}, 193901 (2020).

\bibitem{Farhi2022} A. Farhi, A. Mekawy, A. Al\`{u}, and D. Stone, Excitation of absorbing exceptional points in the time domain, {\pra} {\bf106}, L031503 (2022).

\bibitem{Radi2020} Y. Ra’di, A. Krasnok, and A. Al\`{u}, Virtual critical coupling, {ACS Photon.} {\bf7}, 1468 (2020).

\bibitem{Lepeshov2020} S. Lepeshov and A. Krasnok, Virtual optical pulling force, {Optica} {\bf7}, 1024 (2020).

\bibitem{Ali2021} R. Ali, Lighting of a monochromatic scatterer with virtual gain, {Phys. Scr.} {\bf96}, 095501 (2021).

\bibitem{Kim2022} S. Kim, S. Lepeshov, A. Krasnok, and A. Al\`{u}, Beyond Bounds on Light Scattering with Complex Frequency Excitations, {\prl} {\bf129}, 203601 (2022).

\bibitem{Gu2022} Z. Gu, H. Gao, H. Xue, J. Li, Z. Su, and J. Zhu, Transient non-Hermitian skin effect, {Nat. Commun.} {\bf13}, 7668 (2022).

\bibitem{AllenBook} L. Allen and J.H. Eberly, \textit{Optical Resonance and Two-Level Atoms}, (Wiley, New York, 1975).

\bibitem{Lamb1971} G. L. Lamb, Analytical Descriptions of Ultrashort Optical Pulse Propagation in a Resonant Medium, {\rmp} {\bf43}, 99 (1971).

\bibitem{Kryukov1970} P. G. Kryukov and V. S. Letokhov, Propagation of a Light pulse in a Resonantly amplifying (absorbing) medium, {Sov. Phys. Usp.} {\bf 12}, 641 (1970).

\bibitem{McCall1969} S. L. McCall and E. L. Hahn, Self-Induced Transparency, {Phys. Rev.} {\bf183}, 457 (1969).

\bibitem{Poluektov1975} I. A. Poluektov, Yu. M. Popov, and V. S. Roitberg, Self-induced transparency effect, {Sov. Phys. Usp.} {\bf 17}, 673 (1975).

\bibitem{Crisp1970} M. D. Crisp, Propagation of Small-Area Pulses of Coherent Light through a Resonant Medium, {\pra} {\bf1}, 1604 (1970).

\bibitem{Ziolkowski1995} R. W. Ziolkowski, J. M. Arnold, and D. M. Gogny, Ultrafast pulse interactions with two-level atoms, {\pra} {\bf52}, 3082 (1995).

\bibitem{Kalosha1999} V. P. Kalosha and J. Herrmann, Formation of Optical Subcycle Pulses and Full Maxwell-Bloch Solitary Waves by Coherent Propagation Effects, {\prl} {\bf83}, 544 (1999).

\bibitem{Tarasishin2001} A. V. Tarasishin, S. A. Magnitskii, V. A. Shuvaev, and A. M. Zheltikov, Evolution of ultrashort light pulses in a two-level medium visualized with the finite-difference time domain technique, {Opt. Express} {\bf8}, 452 (2001).

\bibitem{Novitsky2012} D. V. Novitsky, Propagation of subcycle pulses in a two-level medium: Area-theorem breakdown and pulse shape, {\pra} {\bf86}, 063835 (2012).

\bibitem{Arkhipov2020} R. M. Arkhipov, M. V. Arkhipov, and N. N. Rosanov, Unipolar light: existence, generation, propagation, and impact on microobjects, {Quant. Electron.} {\bf50}, 801 (2020).

\bibitem{Afanas'ev1990} A. A. Afanas'ev, V. M. Volkov, V. M. Dritz, and B .A. Samson, Interaction of Counter-propagating Self-induced Transparency Solitons, {\jmo} {\bf37}, 165 (1990).

\bibitem{Shaw1991} M. J. Shaw and B. W. Shore, Collisions of counterpropagating  optical solitons, {\josab} {\bf8}, 1127 (1991).

\bibitem{Arkhipov2021} R. M. Arkhipov, Electromagnetically Induced Gratings Created by Few-Cycle Light Pulses (Brief Review), {JETP Lett.} {\bf113}, 611 (2021).

\bibitem{Novitsky2012a} D. V. Novitsky, Controlled absorption and all-optical diode action due to collisions of self-induced-transparency solitons, {\pra} {\bf85}, 043813 (2012).

\bibitem{Afanas'ev2002} A. A. Afanas'ev, R. A. Vlasov, O. K. Khasanov, T. V. Smirnova, and O. M. Fedotova, Coherent and incoherent solitons of self-induced transparency in dense, resonant media, \josab {\bf19}, 911 (2002).

\bibitem{Ponomarenko2010} S. A. Ponomarenko and S. Haghgoo, Self-similarity and optical kinks in resonant nonlinear media, {\pra} {\bf82}, 051801(R) (2010).

\bibitem{Novitsky2017} D. V. Novitsky, Optical kinks and kink-kink and kink-pulse interactions in resonant two-level media, {\pra} {\bf95}, 053846 (2017).

\bibitem{Novitsky2018} D. V. Novitsky, Disordered resonant media: Self-induced transparency versus light localization, {\pra} {\bf97}, 013826 (2018).

\bibitem{Novitsky2021} D. V. Novitsky, D. Lyakhov, D. Michels, D. Redka, A. A. Pavlov, and A. S. Shalin, Controlling wave fronts with tunable disordered non-Hermitian multilayers, {Sci. Rep.} {\bf11}, 4790 (2021).

\bibitem{Asselie2022} S. Asselie, A. Cipris, and W. Guerin, Optical interpretation of linear-optics superradiance and subradiance, {\pra} {\bf106}, 063712 (2022).

\bibitem{Novitsky2009} D.V. Novitsky, Compression of an intensive light pulse in photonic-band-gap structures with a dense resonant medium, {\pra} {\bf79}, 023828 (2009).

\bibitem{Hughes1998} S. Hughes, Breakdown of the Area Theorem: Carrier-Wave Rabi Flopping of Femtosecond Optical Pulses, {\prl} {\bf81}, 3363 (1998).


\end{thebibliography}
\end{document}